\documentclass{article}
\usepackage[letterpaper, margin=0.75in]{geometry}
\usepackage{authblk}
\usepackage[utf8]{inputenc}
\usepackage{polski}
\usepackage{graphicx}
\usepackage[english]{babel}
\usepackage{amsfonts}
\usepackage{xcolor}
\usepackage{parskip}
\usepackage{float}
\usepackage{hyperref}
\usepackage{wrapfig}
\usepackage{caption}
\usepackage{subcaption}
\usepackage[normalem]{ulem}
\usepackage{amsmath}
\usepackage{cite}
\usepackage{lineno}

\title{Detection of light emission produced in the process of positronium formation}
\author[1]{M.~Pietrow\thanks{corresponding author; email: marek.pietrow@umcs.pl}}
\author[1]{R.~Zaleski}
\author[2]{A.~Wagner}
\author[3]{P.~S{\l}omski}
\author[2]{E.~Hirschmann}
\author[4]{R.~Krause-Rehberg}
\author[2]{M.O.~Liedke}
\author[2]{M.~Butterling}
\author[2]{D.~Weinberger}
\affil[1]{Institute of Physics, M. Curie-Sk{\l}odowska University, ul. Pl. M. Curie-Sk{\l}odowskiej 1, 20-031 Lublin, Poland}
\affil[2]{Institut f\"ur Strahlenphysik / Institute of Radiation Physics, Helmholtz-Zentrum Dresden-Rossendorf, Bautzner Landstr.~400, 01328 Dresden, Germany}
\affil[3]{IT Company \emph{Martinex}, ul. Me{\l}giewska~95, 21-040 {\'S}widnik, Poland}
\affil[4]{Institut f\"ur Physik, Universit\"at Halle, von-Danckelmann-Platz~3, 06120 Halle, Germany}

\date{\today}
\begin{document}
\renewcommand{\tablename}{TABLE}
\renewcommand{\figurename}{FIG.}
\renewcommand{\refname}{References}
\maketitle
\begin{figure}
\centering
\includegraphics[width=8.6cm]{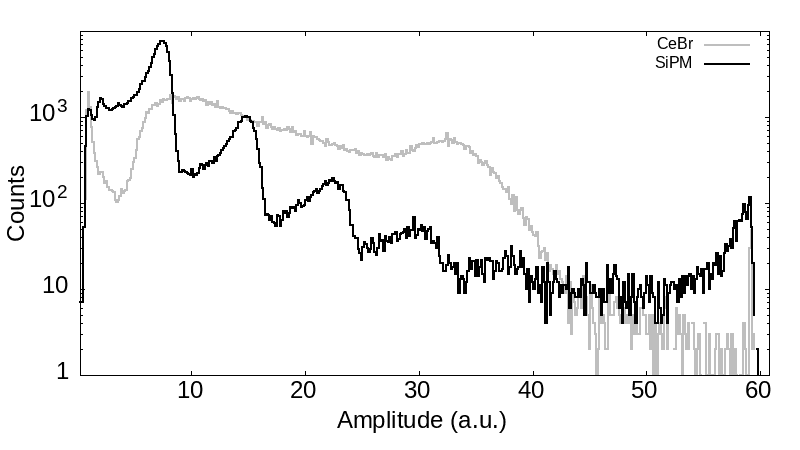}
\caption{Histogram of signal amplitudes from the annihilation radiation detector (CeBr$_3$) -- gray and the UV-Vis light detector (SiPM) -- black.}
\label{fig:AmplitudeSpectrum}
\end{figure}
\begin{figure}
\centering
\includegraphics[width=8.6cm]{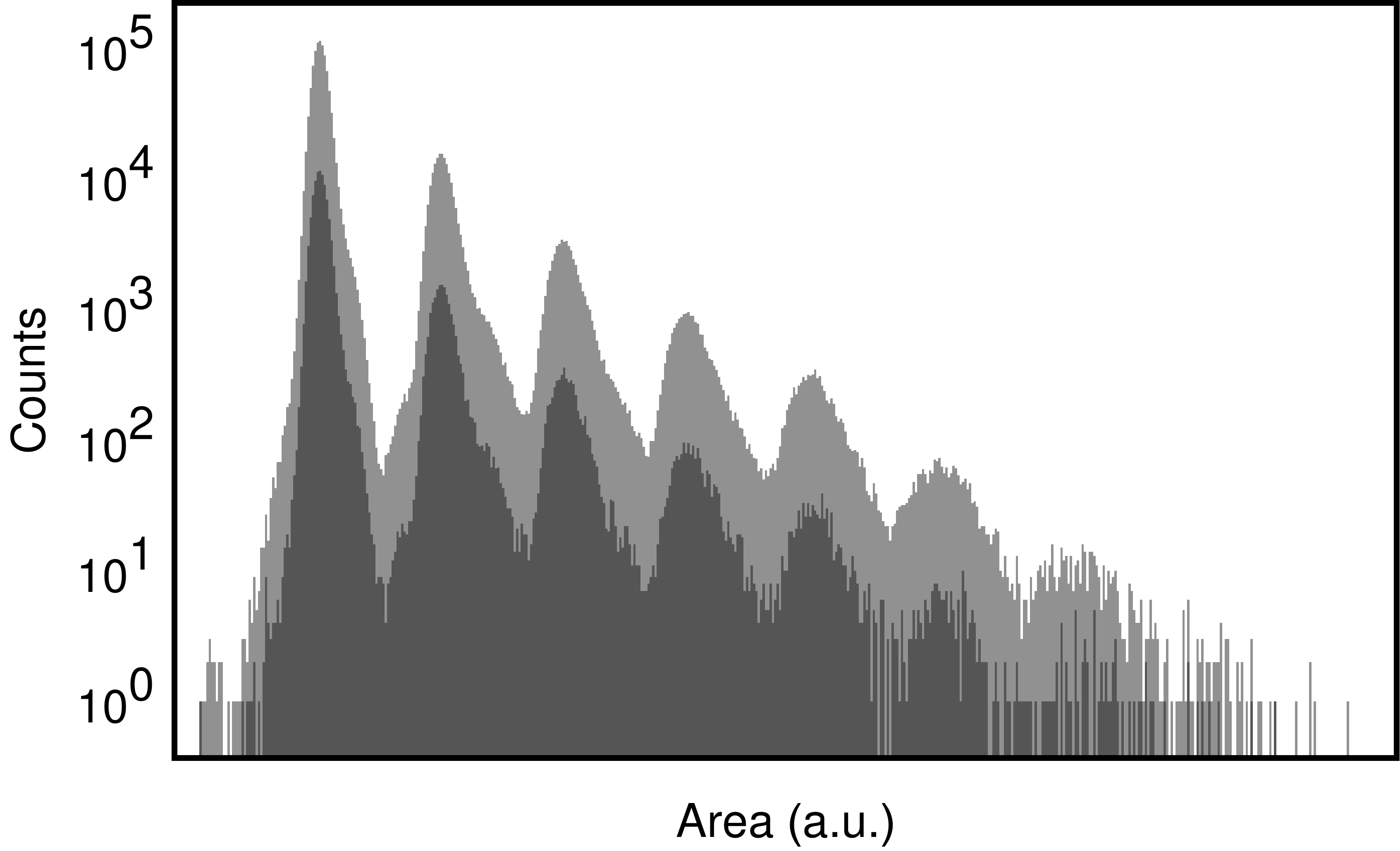}
\caption{Histogram of pulse areas detected in the SiPM channel for n-octacosane collected from photons accompanying all positrons (light shade) and only positrons annihilating after 2~ns from their formation (dark shade).}
\label{fig:NumbersOfPhotons}
\end{figure}
\begin{figure}
\centering
\begin{subfigure}{0.49\textwidth}
\centering
\includegraphics[width=8.6cm]{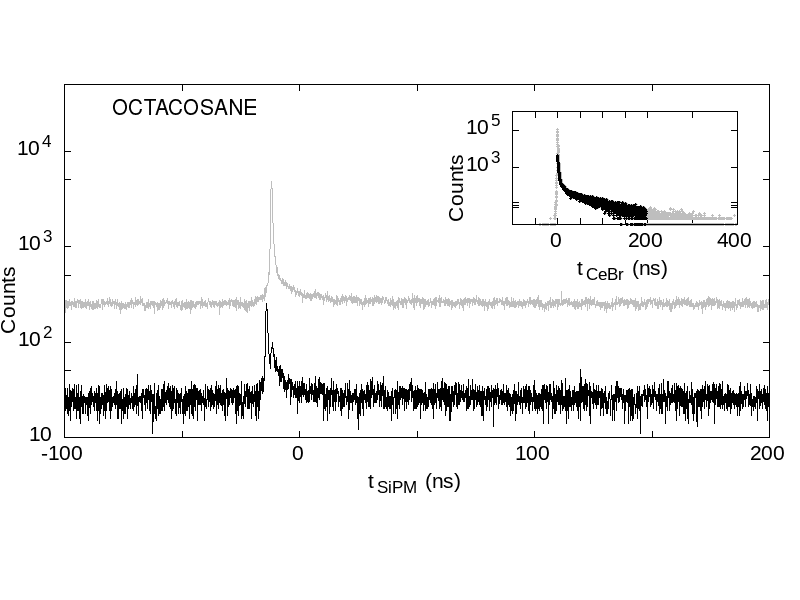}
\end{subfigure}
\begin{subfigure}{0.49\textwidth}
\centering
\includegraphics[width=8.6cm]{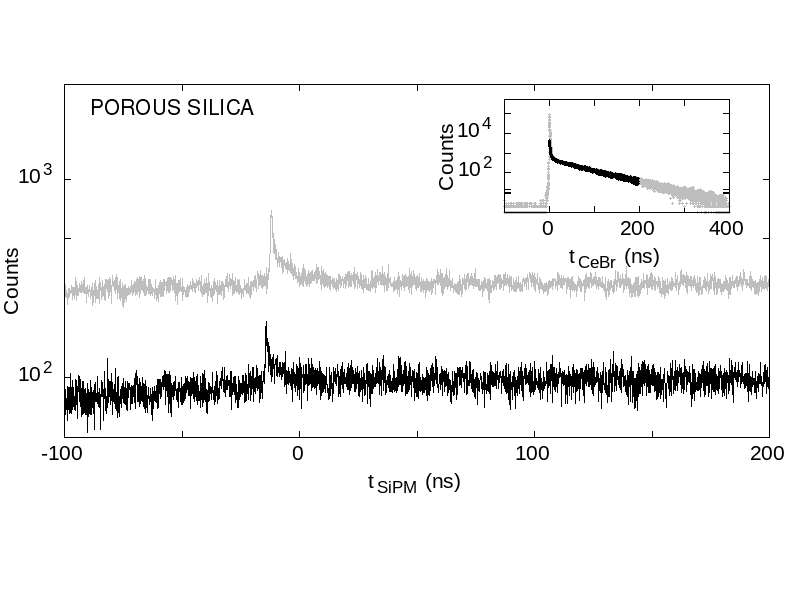}
\end{subfigure}
\caption{Light time of emission spectra for photons accompanying all positrons (gray) and the long-lived Ps fraction (black) for n-octacosane (left) and silica (right). The time histograms of annihilation quanta (positron annihilation lifetime spectra) are shown in respective colors the insets.}
\label{fig:Correlations}
\end{figure}
\begin{figure}
\centering
\begin{subfigure}{0.49\textwidth}
\centering
\includegraphics[width=8.6cm]{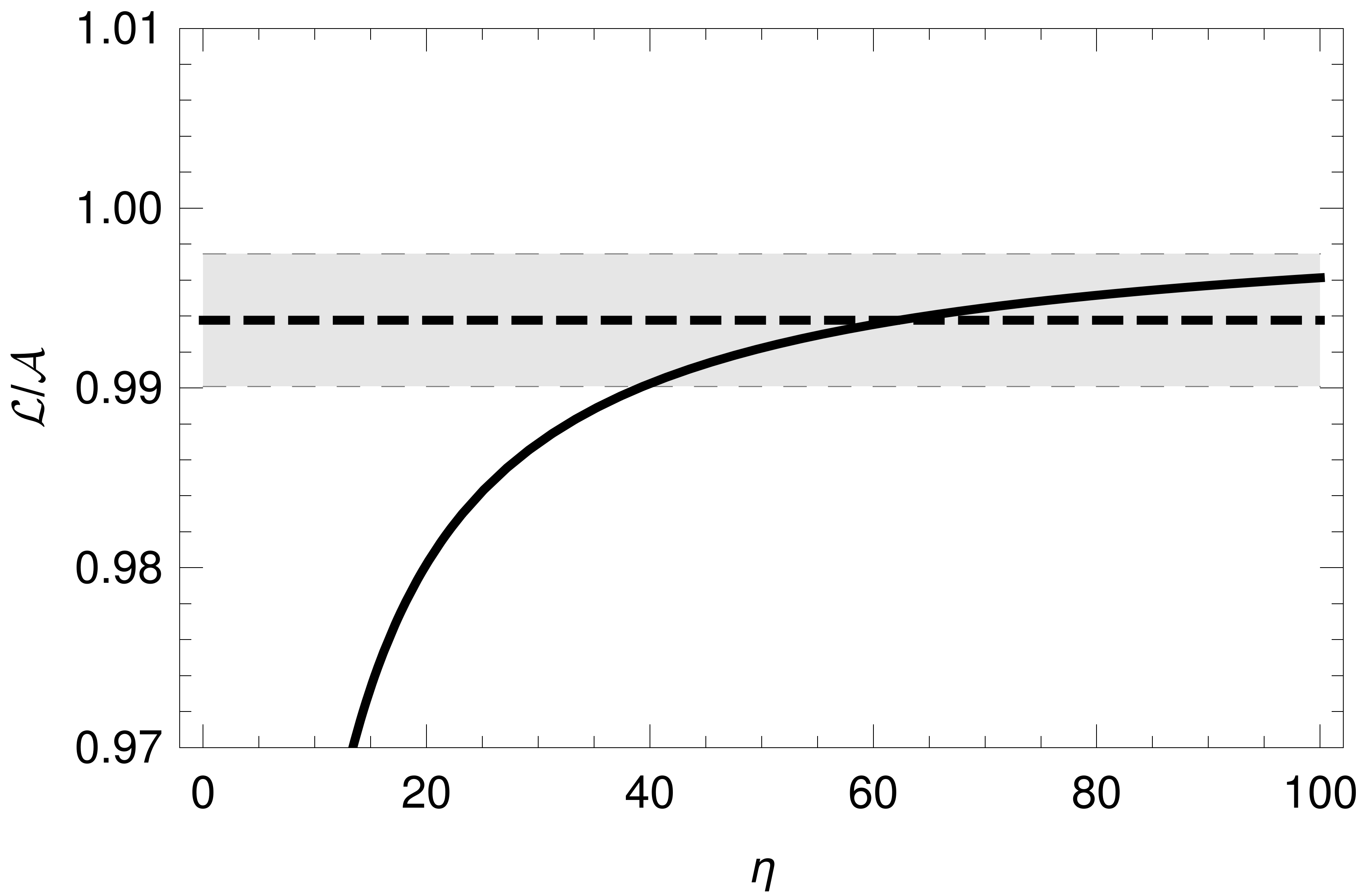}
\end{subfigure}
\begin{subfigure}{0.49\textwidth}
\centering
\includegraphics[width=8.6cm]{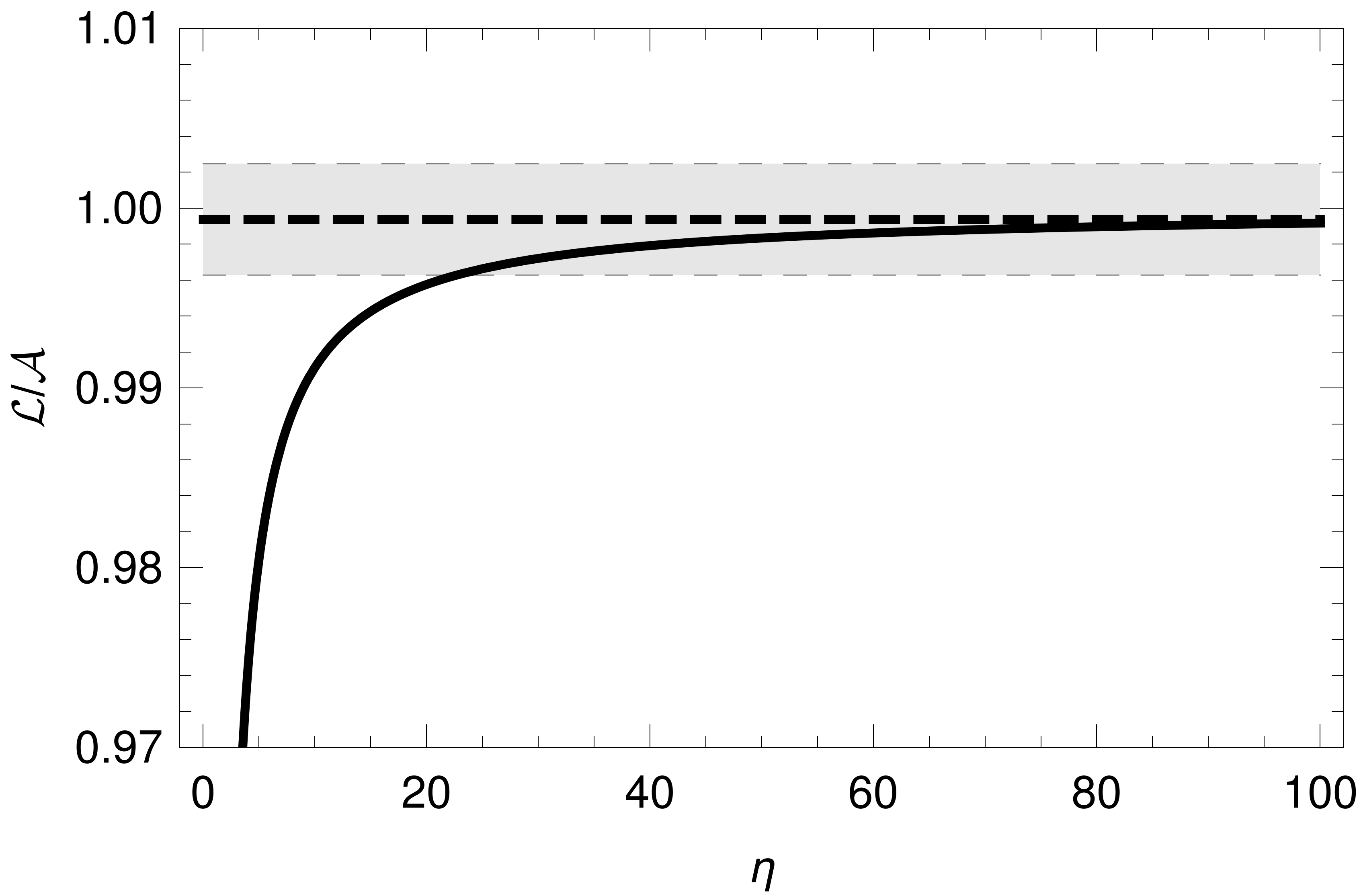}
\end{subfigure}
\caption{Expected value of $\mathcal {L}/\mathcal{A}$ for n-octacosane (left) and silica (right) as a function of the number of background photons $\eta$ (solid line), assuming that the probability of producing an extra photon given by the formulas (\ref{eq:Expect2}) is $\delta$=0. The bold dashed line and the shaded area indicate the value obtained in our experiment, including the uncertainty range.}
\label{fig:LtoA_approx}
\end{figure}

%
\textbf{Abstract:}
The excess energy emitted during the positronium (Ps) formation in condensed matter may be released as light. Spectroscopic analysis of this light can be a new method of studying the electronic properties of materials. We report the first experimental attempt, according to our knowledge, to verify the existence of this emission process. As a result, the possibility of the emission of photons during Ps formation is within the experimental uncertainty in two different solids: an n-alkane and porous silica. However, it seems that the Ps formation on the alkane surface is not accompanied by the emission of photons with energy in the detection range of 1.6 -- 3.9~eV. Various processes that can influence the energy of the photon emitted during the Ps formation are discussed to elucidate this issue. To aid future experiments, equations were developed to estimate the expected ratio of light emission events to annihilation events with the presence or absence of a photon during the Ps formation.
\\
\textbf{Introduction:}
Production of exotic atoms is a task undertaken as part of basic research, e.g. to verify physical theories~\cite{Cassidy07,Acin01,Hiesmayr17,Mohammed19}. These atoms are also used as tools of contemporary applied physics. One of the exotic atoms is the Ps, whose properties in vacuum are quite well known~\cite{Karshenboim04,Rubbia04}. However, its complex behavior inside matter is still discussed since it is used as a non-destructive probe for studying the size of free volumes (e.g. inter-molecular spaces or pores)~\cite{Goworek15}. Furthermore, the positron annihilation phenomenon allows functioning of Positron Emission Tomography which is one of the most influential techniques in medical diagnostics~\cite{Jones17}.
\\
In this paper, we present the results of the experimental verification of the generation of photonic luminescence during the Ps formation in condensed matter. This phenomenon is closely associated with studies of the controlled emission of single photons used in optoelectronics~\cite{Linhart19}. The experiment is supported by previous theoretical calculations of the energy range of light quanta expected to be emitted in this process in alkanes~\cite{Pietrow11,Pietrow17}.
\\
In many investigations, the Ps in a solid is regarded to be an atom identical to that in vacuum. It is formed from a quasi-free $e^+$ -- $e^-$ pair that localizes in a free volume~\cite{Stepanov03}. However, the time evolution of the Ps state in a solid is a complex issue where estimation of the energy thereof should be done with care. Especially, the interaction of the $e^+$ -- $e^-$ pair near the free volume should be discussed as a many-body problem. Due to the interaction screening (caused by permanent dipole moments or by polarizability), the constituents of this pair are loosely bound when they diffuse through the bulk toward the free volume. In turn, their interaction with the surrounding molecules may be described as a trapping process~\cite{Pietrow11}. Furthermore, the similarity of the Ps in a free volume to the vacuum Ps depends on coupling strength with surroundings. In fact, the Ps bond energy is modified due to the interaction of its constituents with the wall of a free volume with a sufficiently small size~\cite{Marlotti16,Pietrow17}. The interaction with the wall should not be neglected for radii of free volumes smaller than about 5$r_{Ps}$ ($r_{Ps}$ -- the effective Ps Bohr radius).\\
The formation of the Ps state is a spontaneous (energetically profitable) process for most substances. When the $e^+$ -- $e^-$ pair enters the free volume
the Ps starts to exist as a bound system in the potential well
and an excess energy is released~\cite{Pietrow17}. This energy can be released as light (UV and Vis region, e.g.: about 3.5~eV for typical alkane~\cite{Pietrow11}). Obviously, the energy of the emitted quanta is in relation to the binding energy of electrons in the sample. 
\\
In general, the photonic deexcitations compete with the phononic ones~\cite{Puska94,Perkins70} but the probabilities of these processes depend on the sample properties. The possibility of phononic deexcitations is not obvious in the molecular crystals of alkanes due
to the low value of the melting transition enthalpy which is about 40kJ/mol for a typical alkane~\cite{MeltingEnthalpy}. This
implies that the estimated energy can cause local melting when it dissipates into the surrounding of a free volume (imparting it with a few molecules forming the wall of a free volume). However, there is no experimental confirmation of such a phenomenon at this time.\\
In practice, the presumed emission of light quanta related to the Ps formation is always accompanied by the emission of near-UV light due to the sample irradiation by positrons~\cite{Mozumder99,MozumderHatano03}. Positrons are usually available from a $\beta^+$ radioisotope or from positrons generated via pair production, especially delivered in a positron beam. In both cases they lose the kinetic energy (of the order from a few to hundreds of keV) producing ion-electron pairs and orbital electronic excitations. Light is emitted as a consequence of deexcitations of these species. Additionally, Cherenkov radiation is emitted by positrons or electrons with kinetic energy greater than about 0.2~MeV~\cite{KonyaNagy12} for substances with a refractive index characteristic for alkanes and silica. Furthermore, recently, intensified luminescence (compared to that stimulated by electrons with the same energy) of $\beta^+$ irradiated samples has been observed~\cite{PsLuminescence18}. Additional light is emitted due to the Auger process, which is related to the positron annihilation with a core electron.
\\
Our attempt to observe the emission of light during the Ps formation gives new insight into the physical processes accompanying positronium formation in condensed matter. Additionally, the spectroscopy of light emitted during Ps formation should be able to provide information about the electronic properties of condensed matter. In this way, the method would extend the capabilities of the classical positron annihilation spectroscopy.\\
\\
\textbf{Experimental setup:}
Prior to the beam experiment, positron annihilation lifetime spectroscopic (PALS) measurements were performed using a $^{22}$Na radioisotope source (0.2~MBq). Two scintillation detectors equipped with BaF$_2$ scintillators and Hamamatsu H3378-50 assemblies were coupled with a two channel Agilent U1065A digitizer with a sampling rate of 4~GS/s and 10-bit resolution. The digitizer was triggered by a custom-made fast coincidence unit. The digitized impulses were analysed~\cite{Prague} in-flight providing positron lifetime spectra. The samples were placed in the sandwich geometry in the vacuum chamber (pressure below 10$^{-4}$~Pa) at room temperature. Over 20M counts were collected.\\
Subsequently, to avoid a high light background produced by energetic positrons, the MePS positron beam at ELBE was used as a source of slow positrons~\cite{Wagner18}. The energy of the electron beam was 33.2~MeV, its intensity -- 98~pA, and the repetition rate -- 1.625~MHz with a 615~ns bunch spacing. It produced a 2~keV positron beam.\\
The annihilation radiation was detected using a CeBr$_3$ scintillator coupled with a photomultiplier. The UV-Vis photons were detected using a 6x6~mm$^2$ Hamamatsu S13360-6075PE silicon photomultiplier (SiPM) with photon detection efficiency above 20\% in the wavelength range of 330--800~nm (1.6--3.8~eV) and a peak sensitivity at 450~nm. It was mounted on a fluid cooled preamplifier board fitted on a stainless steel flange.
\\
Cooling of the flange with a light detector and the sample was supplied by a Fryka DLK 632 cooling unit with HKF 15.1 fluid operated at -12$^\circ$C, which resulted in a temperature of the sample of about -2$^\circ$C.
\\
Signals from the detectors were collected by an Agilent U1065A 4 channel digitizer with 10-bit resolution. Three channels were used for registration -- the logical signal for electron beam pluse time reference (referred as RF), the analog signal for annihilation events (CeBr$_3$), and the analog signal for photon events (SiPM). Triggering with the SiPM signal resulted in about 640k waveforms (for each of all 3 channels) per hour. All three signals were digitized with the sampling frequency of 2~GS/s (0.5~ns sample interval) and a waveform time length of 800~ns. 1.5M events were recorded for each sample. The Dynamic Range (DR) was 40 for the signal in the CeBr$_3$ channel and 105 for RF. Thanks to the cooling system, the DR value for the photon detector channel was 20. The collected data were analyzed by a custom-made software written in \texttt{C++}.
\\
The samples had a form of 0.5--1~mm plates fixed with carbon tape on their edges to the active area of the light detector. Two different samples were used. The first one was n-octacosane which forms a polycrystalline molecular structure and is semi-translucent in the Vis region. In this sample, the cavities where the Ps is formed, are mainly inter-lamellar gaps with a thickness less than 0.2~nm~\cite{Goworek15} and volumes formed by the bent parts of molecules forming non-planar conformations (\emph{kink} and \emph{gauche} conformers). The second sample was porous silica with a typical pore diameter of about 4~nm~\cite{Sienkiewicz17}. The silica sample is almost fully translucent in the Vis region. The samples differ in both physical and chemical properties. The advantage of using the alkane and silica samples is that the probability of the Ps formation is relatively high in both.\\
\\
\textbf{Data analysis:}
%
The amplitudes of CeBr$_3$ signals (fig.~\ref{fig:AmplitudeSpectrum}) are related to the energy deposited in the detector by a single annihilation quantum. Differently, detection of each light photon in the SiPM detector is followed by formation of a signal with nearly the same amplitude and a shape and, consequently, the same area.
Many of the multi-photon SiPM signals are separated in time by tenths of ns and easily distinguished. However, there are often several signals arriving almost simultaneously. Their summed amplitude gives information about the number of light quanta instead of their energy (fig.~\ref{fig:AmplitudeSpectrum}).
The number of all photons forming the SiPM signals is recognizable (even for many overlapping and shifted in time pulses) based on the area under the pulse curve (fig.~\ref{fig:NumbersOfPhotons}).
\\
To approximate the pulse arrival time, we used the Levenberg-Marquardt method to fit the data~\cite{NumericalRecipes} for both the CeBr$_3$ and SiPM pulses (additionally, some modifications of the fitting algorithm as trapezoidal data filtering were tested). As the fitting function, we used
\begin{equation}
\frac{I}{2}\ [Y(t_k;\tau,s)-Y(t_{k+1};\tau,s)+\Phi(t_k/s)-\Phi(t_{k+1}/s)],
\label{eq:LTpart1}
\end{equation}
where
\begin{equation}
Y(t;\tau,s)= e^{\frac{s^2}{4\tau^2}} \ \Phi \Big( \frac{s}{2\tau}-\frac{t}{s}\Big) \ e^{-\frac{t}{\tau}}, \ \Phi(x)=\textrm{erfc}(x).
\label{eq:LTpart2}
\end{equation}
The fitting function is a convolution of an exponential function with a Gaussian,
where $t_k$ -- time related to the $k$-th sample in the waveform, whereas $I$, $\tau$, and $s$ are fitting parameters describing the amplitude and the slope of the fitting curve.
This yields a histogram of time differences between RF and CeBr$_3$ or SiPM signals, i.e. time histograms of annihilation quanta (positron annihilation lifetime spectra) and time of emission spectra for light photons.\\
\\
\textbf{Results:}
Table~\ref{tab:Lifetimes} shows the results of the decomposition of the positron lifetime spectra (CeBr$_3$ channel) into components using the LT program~\cite{Kansy96} for both samples. They are compared to the results obtained for these samples using the $^{22}$Na radioisotope as a positron source.
\begin{table*}
\scriptsize
\centering
\begin{tabular}{|c c c c c c c c c c c|}
\hline
Sample/(positron source) & $I_1$ (\%) & $\tau_1$ (ns) & $I_2$ (\%) & $\tau_2$ (ns) & $I_3$ (\%) & $\tau_3$ (ns) & $I_4$ (\%) & $\tau_4$ (ns) & $I_5$ (\%) & $\tau_5$ (ns)\\
\hline\hline
n-octacosane (beam)& 35.1(2) & 0.1(1) & 41.4(2) & 0.4(1) & 18.2(1) & 1.4(2) & 1.9(4) & 7.2(6) & 3.3(4) & 57.9(9)\\
n-octacosane ($^{22}$Na) & 31.0(4) & 0.2(1) & 42.0(4) & 0.35(5) & 27.0(4) & 1.4(1) & - & - & - & -\\
\hline
silica (beam) & 51.2(7) & 0.3(1) & 17.3(7) & 1.1(2) & 2.4(4) & 5.4(9) & 1.6(6) & 27(3) & 27.5(7) & 80(1)\\
silica ($^{22}$Na) & 52.4(3) & 0.17(5) & 22.5(2) & 0.55(5) & 2.2(2) & 3.6(1) & 1.7(7) & 24.4(3) & 21.2(6) & 51.0(2)\\
\hline
\end{tabular}
\caption{Lifetimes and intensities for n-octacosane and silica samples obtained with the slow positron beam and the $^{22}$Na radioisotope.
}
\label{tab:Lifetimes}
\end{table*}
%
\\
In the case of both samples, the first component ($\tau_1$, $I_1$) is related to the para-Ps fraction (a short bulk e$^+$ component is possible) whereas the second one ($\tau_2$, $I_2$) is related to annihilation of $e^+$ without forming a bound state. In the case of the beam experiment, the parameters of both these components are determined with very low accuracy due to the relatively wide resolution function at such low positron energy.
The lifetimes of the other components ($\tau_3$ -- $\tau_5$) exceed 1~ns and are more easily resolved. They are related to several fractions of the ortho-Ps located in cavities of different size. In the alkane, the $\tau_3$=1.4~ns component is well known form the PALS; it is interpreted as the ortho-Ps annihilation in the interlamellar gaps. However, the respective intensity $I_3$ is much smaller in the experiment with the positron beam. Additionally, some additional long-lived components which are unobserved with the $^{22}$Na radioisotope, appear in the spectrum obtained with the beam. This discrepancy can be explained assuming that, due to the small energy of incident positrons from the beam, some fraction of the Ps is formed on the surface of the sample or very close to it. Then, the Ps can migrate outside the sample and annihilate in vacuum with a lifetime of 142~ns, which is underestimated in the experiment due to the Ps movement away from the detector. Additionally, some small fraction of positrons can be scattered from the sample surface and annihilate outside in the chamber housing. They contribute to the spectra depending on their time of flight.
\\
In the case of the silica, the discrepancies between the results of the measurements with the radioisotope and the beam positron source are also caused by the small implantation range of slow positrons. The dominant Ps fraction is formed in mesopores of the silica. If the radioisotope source is used this fraction also annihilates in mesopores with a characteristic lifetime $\tau_5$ of about 51~ns. However, if the positrons come from the low-energy beam the lifetime of this component becomes 80~ns which suggests that it is related to Ps or $e^+$ annihilating outside the sample. Assuming this~\cite{Gidley06}, the larger lifetime for the silica than for the alkane suggests that the velocity of the migrating Ps is smaller in the former case. This is reasonable, because Ps leaving pores can lose part of its energy while scattered on pore walls. Alternatively, from the quantum-mechanical point of view, Ps has zero-point energy, which is converted into kinetic energy, which is lower in larger cavities (i.e. mesopores).\\
The components with lifetimes $\tau_3$ and $\tau_4$ are related to the annihilation in micropores. Their parameters are difficult to distinguish precisely from of the long-lived fifth component with high intensity even in the case of the radioisotope experiment due to their low intensity and the statistical dispersion.\\
To select events related to the Ps formation only, the information from the positron annihilation lifetime spectra was used (insets in fig.~\ref{fig:Correlations}). The free positron lifetimes are about 0.3~ns in the n-octacosane and 0.5~ns in the silica, thus only the Ps can survive longer than several nanoseconds. Therefore, the choice of events with annihilation events occurring above some time threshold assures that they are related only to the Ps formation and annihilation. This time range, i.e.~from $t_1$=2~ns to $t_2$=200~ns, was chosen arbitrarily from the possibly smallest value that assures negligible contribution from free positrons and to minimize statistical dispersion of the results.\\
The photon time-of-emission spectra for photons accompanying any positron (both unbound and bound in the Ps) and these associated with the long-lived Ps fraction only are shown for the n-octacosane and the silica in fig.~\ref{fig:Correlations}. The presented spectra are limited to events with a single peak in the SiPM channel, which allowed us to find the time with good accuracy.
For both samples, the photon time-of-emission spectra have a peak whose position correlates with the time of positron implantation. Furthermore, the wide region of approximately homogeneously distributed uncorrelated events (preceding and following the positron implantation time) forms a plateau visible in the figure on the left and on the right from the peak.\\
To determine if there are additional light photons related to the Ps formation, two count ratios are defined, $\mathcal{L}=L_{Ps}/L_{e^+}$, where: $L_{Ps}$ and $L_{e^+}$ are numbers of light photons related to the selected long-lived Ps fraction and to all positrons, respectively, and $\mathcal{A}=A_{Ps}/A_{e^+}$, where: $A_{Ps}$ and $A_{e^+}$ are numbers of annihilation quanta related to the same long-lived Ps fraction and to all positrons, respectively.\\
The $A_{e^+}$ and $A_{Ps}$ can be obtained from the positron annihilation lifetime spectra by summing all counts and the counts in the time region 2--200~ns, respectively. The $L_{e^+}$ and $L_{Ps}$ are calculated as proportional to the area under photonic signals in all events and events accounted in $A_{Ps}$, respectively. The amplitude range of the CeBr$_3$ signals that were taken into account included the 511~keV peak and the Compton scattered $\gamma$ events whereas the amplitude range of the SiPM signals spreads from the single-photon to the four-photon events (see fig.~\ref{fig:AmplitudeSpectrum}).
%
%
Applying the data from our experiment, the $\mathcal{L}$ and $\mathcal{A}$ ratios are $\mathcal{A}$=0.09479(5), $\mathcal{L}$=0.0942(3) for the n-octacosane and $\mathcal{A}$=0.3227(5), $\mathcal{L}$=0.3225(5) for the porous silica which give the $\mathcal{L}/\mathcal{A}$ ratio of 0.994$\pm$0.001 for the n-octacosane and 0.999$\pm$0.001 for the porous silica. The result does not change substantially when the range of the CeBr$_3$ amplitudes is restricted (e.g. taking annihilations from the region of 511~keV peak only), which proves that the different detection efficiency of two and three photon annihilation does not influence the result. Also, changing the SiPM amplitude (e.g. taking one-photon events) or the accepted photonic emission time range does not influence the results significantly.\\
\\
\textbf{Discussion:}
The coefficients $\mathcal{A}$ and $\mathcal{L}$ can be estimated parametrically using obtained annihilation intensities and lifetimes (tab.~\ref{tab:Lifetimes}) in the equation
\begin{equation}
\mathcal{A}=\frac{\sum_{i=1}^{5}I_i (e^{-t_1/\tau_i}-e^{-t_2/\tau_i})}{\sum_{i}I_i}
\label{eq:Expect1}
\end{equation}
which gives $\mathcal{A}$=0.10(1) for the n-octacosane and 0.31(1) for the silica.\\
To estimate the value of $\mathcal{L}$, it was assumed that $\eta_i$ are mean values of photons produced by the $i$-th fraction of energetic positrons during their thermalization whereas $\delta_i$ is the number of extra photons produced per single Ps formation. Then
\begin{equation}
\mathcal{L}= \frac{\sum_{i=1}^{5}I_i (\eta_i+\delta_i) (e^{-t_1/\tau_i}-e^{-t_2/\tau_i})}{\sum_{i}(\eta_i+\delta_i) I_i}.
\label{eq:Expect2}
\end{equation}
It can be assumed that, for any fraction of the free e$^+$, $\eta_i$ has the same value of $\eta$. However, in the case of the Ps, one of the quasi-free electrons is taken for the Ps formation and does not produce a photon. The value of this expression vary with changes in $\eta$ and $\delta$.\\
The simplest case is when no additional photons are produced during the Ps formation. With reference to the constituents shown in~tab.~\ref{tab:Lifetimes} and additionally assuming a lack of additional photon production during the Ps formation, we set $\delta_i=0$ for all $i$-constituents and $\eta_2$=$\eta$ and $\eta_i=\eta-1$ otherwise. The $\eta$ parameter may vary from tens to a few hundreds. In particular, $\mathcal{L}$=0.09(1) for the octacosane and 0.31(1) for the silica if $\eta$=20. In general, $\mathcal{L}/\mathcal{A}$ is a function of $\eta$ shown in fig.~\ref{fig:LtoA_approx}. The comparison of the experimental results presented in the "Results" section with the $\mathcal{L}/\mathcal{A}(\eta)$ curve allows us to conclude that there is no emission of photons during the formation of the Ps only if the emission of approx. 60 de-excitation photons for the n-octacosane and over 100 for the porous silica occurs. Since a rather lower number of de-excitation photons is expected, it appears that the formation of at least some Ps is accompanied by photon emission.
\\
\begin{figure}
\centering
\begin{subfigure}{0.49\textwidth}
\centering
\includegraphics[width=8.6cm]{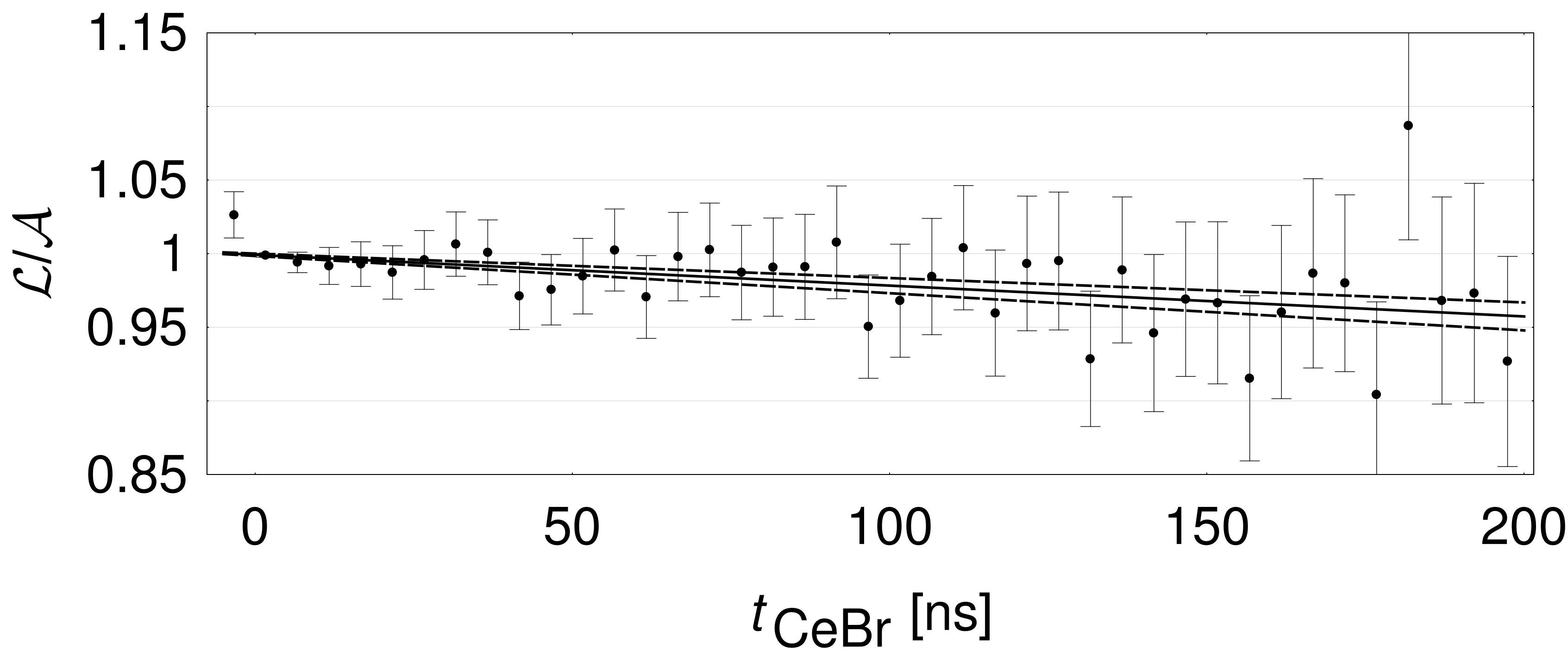}
\end{subfigure}
\begin{subfigure}{0.49\textwidth}
\centering
\includegraphics[width=8.6cm]{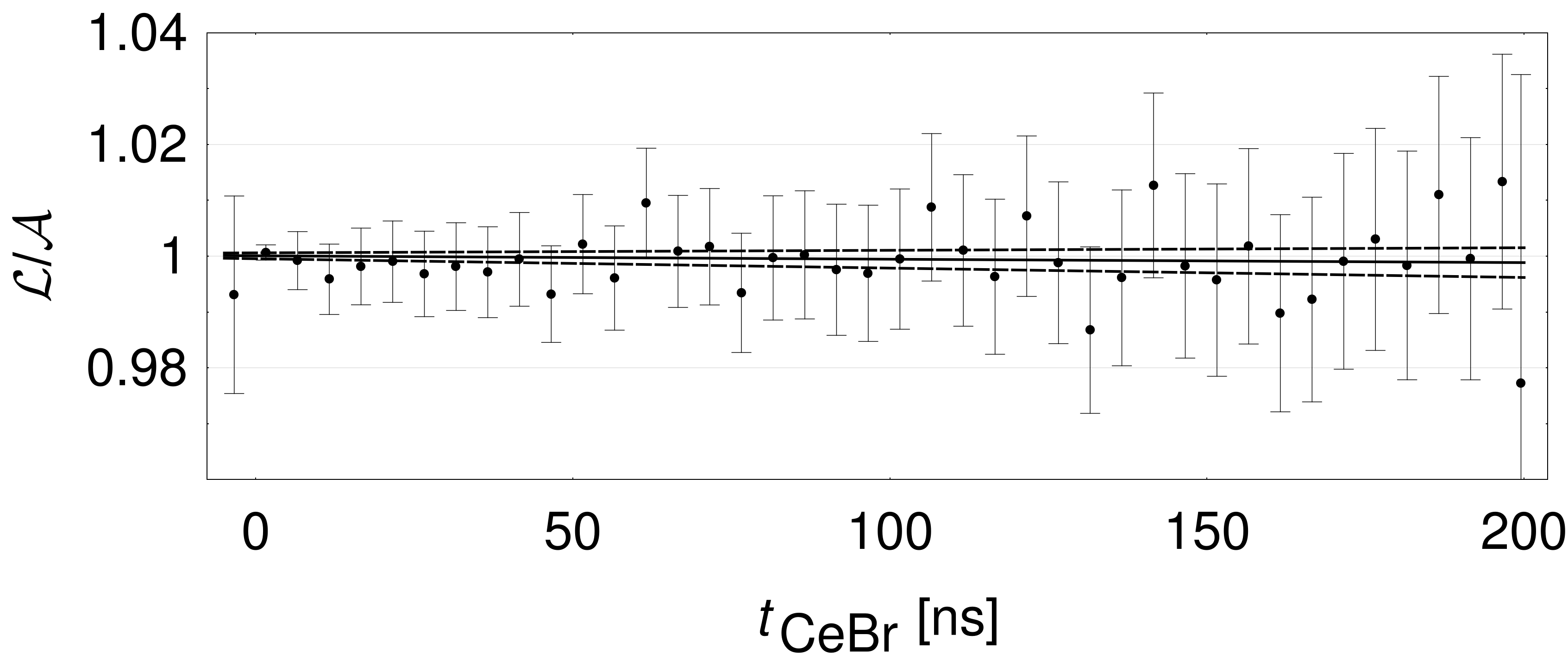}
\end{subfigure}
\caption{$\mathcal {L}/\mathcal{A}$ for n-octacosane and silica, respectively, as a function of time $t_{CeBr_3}$ of the corellated annihilation $\gamma$ emission. The solid line is a least square approximation of this function whereas the dashed lines denote a range of the approximation including standard deviation. The error for each point is indicated by vertical bars.}
\label{fig:LtoA_errors}
\end{figure}
To obtain more detailed information about the $\mathcal{L}/\mathcal{A}$ value for various fractions of annihilating positrons, $\mathcal{L}/\mathcal{A}(t)$ was calculated as a function of the positron annihilation time, i.e. $t_{CeBr_3}$ (fig.~\ref{fig:LtoA_errors}). Linear functions fitted to the experimental data allow estimation of the average $\mathcal{L}/\mathcal{A}$ value as a function of time, despite the quite large scatter of individual experimental points. It is equal to 1 within the experimental uncertainty over the entire time range for the silica (fig.~\ref{fig:LtoA_approx}). Remembering that one of the electrons from the ionization track is involved in the Ps formation and does not produce a deexcitation photon, the absence of photon deficiency suggests that an additional light photon with energy in the detection range is emitted during the Ps formation.\\
The positron implantation profiles for porous silica and polymers, which are representative for our samples, are well known~\cite{Veen03,Algers03,Brusa18}. The median implantation depth is tenths of nm for a 2~keV positron. Consequently, most positrons form Ps in the bulk well below the surface. Nevertheless, in silica, the energetic conditions for the Ps formation on the surface of pores are similar to these on the outer surface of the sample. The equilibrium energy of the formed Ps is slightly higher than in vacuum (i.e. about 0.06~eV more, averaged over occupied energy levels in a 4nm pore) but much lower than in n-octacosane (i.e. even 1.26~eV more for a spherical geometry)~\cite{Shantarovich08}.\\
The scenario of the Ps formation and annihilation in the same cavity is correct for alkanes where the free volumes are well separated and the transition from one free volume to the other is hardly possible. In porous silica, the free volumes are interconnected, which means that Ps can migrate easily from one cavity to another. During the Ps formation, a part of the excess energy is converted into kinetic energy and allows Ps to migrate through the pores. Since the entrances to the pores are located on the silica surface the Ps is able to migrate toward the surface and escape from the sample before its annihilation~\cite{Veen03,Zaleski13}. This process occurs for all Ps formed in mesopores from 2~keV positrons, which confirms the absence of a component with a lifetime specific for pores in the positron annihilation lifetime spectrum for silica. In addition, there is high probability of the Ps formation from the surface state (21\% of implanted positrons) because the internal surface of the pores in silica is very large~\cite{Sienkiewicz17}. The similar history of nearly all positrons that annihilate after a long time explains the lack of $\mathcal{L}/\mathcal{A}$ dependence on time.\\
The systematic decrease in the $\mathcal{L}/\mathcal{A}(t)$ with the positron annihilation time for the n-octacosane suggests a light photon deficiency for long-lived positrons. The comparison of $\mathcal{L}/\mathcal{A}(t)$ with the positron annihilation lifetime spectrum (inset in fig.~\ref{fig:Correlations}) indicates that the photon deficiency concerns mostly low-intensity components with lifetimes of 7.2~ns and 57.9~ns, which are ascribed to positronia formed on the sample surface or close to it. In the time range of 14-200~ns, in which only these components make a significant contribution to the spectrum, the $\mathcal{L}/\mathcal{A}$ is only 0.988$\pm$0.001. Although the positronium cannot migrate in the n-octacosane, a fraction of implanted positrons (about 5\%) can migrate to the surface during thermalization in a diffusive way without forming a bound state with an electron~\cite{Hugenschmidt16}. These positrons can form Ps in the molecule-thick surface layer, in which both the positron and electron are less screened by induced dipoles. The initial state for these particles is more strongly bound and the final state can be a vacuum level, which is over 1~eV lower than the ground level of the Ps in the bulk alkane. Thus, the transition of the e$^+$ -- e$^-$ pair to the Ps state gives less excess energy, which may be outside the detection range.\\
Another process influencing the number of emitted photons is the formation of a bound state of a non-thermalized positron and a molecule, which may be more probable on the sample surface. Consequently, some energy is not transferred to ionizations and excitations, causing photon deficiency. Then, the excess energy of both excitation and Ps formation can be released as a light photon, but the summed energy would be comparable to the energy of the Ps formation in a vacuum (6.8~eV), which is outside the detection energy range.\\
The $\mathcal{L}/\mathcal{A}$ ratio is much closer to 1 (0.996$\pm$0.001) in the time range of 2-14ns, where the contribution from the component related to Ps annihilation in bulk n-octacosane dominates. The difference between the value of $\mathcal{L}/\mathcal{A}$ and 1 agrees very well with the 30\% contribution in this time range of the two longest-lived components for which $\mathcal{L}/\mathcal{A}$=0.988 was calculated previously. Therefore, it can be assumed that $\mathcal{L}/\mathcal{A}$ is nearly 1 for the Ps formed in bulk n-octacosane suggesting the emission of an additional photon with energy in the detection range during the Ps formation.
\\
\\
\textbf{Conclusions:}
Our experiment did not confirm unquestionably an additional light emission related to the Ps in the wide wave length range (about 330-800~nm). However, assuming that one of the electrons from the ionization track is used in the Ps formation, the same number of photons for unbound positrons and positronium suggests that the missing photon from de-excitation is compensated for by the photon emitted during the Ps formation. The formulas~(\ref{eq:Expect1}) and (\ref{eq:Expect2}) presented above allow estimation of the expected ratio of light emission events to annihilation events ($\mathcal{L}/\mathcal{A}$) with the presence and absence of a photon during Ps formation for different positron annihilation lifetimes. This allows choosing a sample with the best possible $\mathcal{L}/\mathcal{A}$ dependency on the number of deexcitation photons and designing an experiment whose results will have uncertainties small enough to determine indisputably if light emission accompanies the Ps formation. In addition, some theoretical considerations should be discussed to help verify the current result in the future.\\
The energy difference between the initial and the final state during the Ps formation may be out of the measured range of 1.6 -- 3.9~eV. The most difficult to calculate and verify is the initial state energy. It was assumed here that the $e^+$ and $e^-$ particles are screened by the dipole field of traps whose structure cannot be changed permanently due to the stable position of molecules in the molecular structure. If this assumption is inaccurate, the increase in the potential energy in the time scale of the $e^+$ -- $e^-$ pair transition into a free volume would be a continuous process and, possibly, the energy emission would not proceed in a stepwise mannner. In this case, a more probable way of the release of the excess energy would be molecular vibrations. Perhaps, the temperature decrease can reduce the dipole relaxation capability and make the stepwise transition more likely.\\
The second issue is related to the final state of the pair. The size of the cavities in the silica is large enough to allow formation inside the vacuum-like Ps. However, for the n-octacosane, the pair should be considered as a confined one and its binding energy would not be as large as in the vacuum atom. In this case, a lesser energy excess is expected. To verify this, changing the detection range of the photon energy towards the infrared region could give a positive result.\\
Furthermore, before entering the free volume, both the $e^+$ and $e^-$ should be regarded as particles in the potential well made by particular molecules. The particles are localized in the local minima of the potential and their escape from one molecule to the other requires positive work. Especially, in the case of alkanes, molecules interact by weak van der Waals interactions. The transition of $e^-$ and $e^+$ toward the free volume from the vicinity of one molecule to the other one becomes difficult and requires work against the field of valence electrons. Therefore, the transition toward the free volume may, in general, require additional energy for transitions through a set of potential steps. In this regard, there is a process of changes in molecule conformations (\emph{kink} and \emph{gauche} conformers~\cite{Maroncelli85}) stimulated by the thermal energy (meV) in alkanes. Additional structural changes of these molecules are possible due to the $e^+$ -- $e^-$ excess energy during Ps formation. These processes diminish the difference between the initial and final state energy during Ps formation and thus the energy of the hypothetical photons. Possibly, a less energy-expensive transition process may be found for other compounds.
\bibliographystyle{ieeetr}
\bibliography{references}
\end{document}